\documentclass[aps,prd,reprint,showpacs,superscriptaddress,twocolumn]{revtex4-2}
\usepackage{amsfonts}
\usepackage{amsmath}
\usepackage{amssymb}
\usepackage{dsfont}
\usepackage{hyperref}
\usepackage{graphicx}
\usepackage{float}
\usepackage[usenames,dvipsnames]{xcolor}
\usepackage[normalem]{ulem}
\usepackage{times}
\usepackage{braket}
\usepackage{booktabs}
\usepackage{multirow}

\hypersetup{
    colorlinks=true,linkcolor=blue,citecolor=blue,
    filecolor=blue,urlcolor=blue,breaklinks=true
}
\RequirePackage{color}

\begin{document}

\date{\today}

\title{Cosmic string influence on a 2D hydrogen atom and its relationship with the Rytova-Keldysh logarithmic approximation in semiconductors}

\author{Frankbelson dos S. Azevedo}
\email{frfisico@gmail.com}
\affiliation{Departamento de F\'{\i}sica, Universidade Federal do Maranh\~{a}o, 65085-580 S\~{a}o Lu\'{\i}s, Maranh\~{a}o, Brazil}

\author{Izael A. Lima}
\email{izael@fisica.ufc.br}
\affiliation{Departamento de F\'{\i}sica, Universidade Federal do Maranh\~{a}o, 65085-580 S\~{a}o Lu\'{\i}s, Maranh\~{a}o, Brazil}

\author{Gallileu Genesis}
\email{galileugenesis@gmail.com}
\affiliation{Departamento de Engenharia Civil, Universidade Federal de Pernambuco,
 50670-901, Recife, Pernambuco, Brazil}
\date{\today}

\author{Rodolfo Casana}
\email{rodolfo.casana@ufma.br}
\affiliation{Departamento de F\'{\i}sica, Universidade Federal do Maranh\~{a}o, 65085-580 S\~{a}o Lu\'{\i}s, Maranh\~{a}o, Brazil}

\author{Edilberto O. Silva}
\email{edilberto.silva@ufma.br}
\affiliation{Departamento de F\'{\i}sica, Universidade Federal do Maranh\~{a}o, 65085-580 S\~{a}o Lu\'{\i}s, Maranh\~{a}o, Brazil}

\begin{abstract}

A two-dimensional hydrogen atom offers a promising alternative for describing the quantum interaction between an electron and a proton in the presence of a straight cosmic string. Reducing the hydrogen atom to two dimensions enhances its suited to capture the cylindrical/conical symmetry associated with the cosmic string, providing a more appropriate description of the physical system. After solving Schr\"dinger's equation, we calculate the eigenenergies, probability distribution function, and expected values for the hydrogen atom with logarithmic potential under the influence of the topological defect. The calculations for the 2D hydrogen atom are performed for the first time using the Finite Difference Method. The results are presented through graphics, tables, and diagrams to elucidate the system's physical properties. We have verified that our calculations agree with a linear variational method result. Our model leads to an interesting analogy with excitons in a two-dimensional monolayer semiconductor located within a specific semiconductor region.
To elucidate this analogy, we present and discuss some interaction potentials and their exciton eigenstates by comparing them with the results from the literature.

\end{abstract}

\maketitle

\section{Introduction}

The non-relativistic and relativistic three-dimensional hydrogen atom studied in curved spacetimes allows us to observe the effects of curvature on its energy spectrum, as noted in Refs. \cite{de2002non, parker1980one}. For instance, the spectrum's behavior of a non-relativistic quantum particle interacting with different potentials in spacetimes generated by topological structures, such as the straight cosmic string and global monopoles, has been evaluated in Refs. \cite{falkenberghydrogen, marques2002hydrogen, marques2007some}. Particularly, in Ref. \cite{de2002non}, a quantum particle interacting with a Coulomb potential in the presence of a straight cosmic string of the Grand Unified Theory (GUT) was studied, revealing that the energy spectra differ from the Euclidean values by $4 \times 10^{-3}$. Specifically, cosmic strings are topological defects that may have been formed during spontaneous symmetry breaking in the early universe. Even if they are not yet experimentally verified, they are well-predicted objects in GUT's realm. Besides, cosmic strings are analogous to defects found in condensed matter \cite{moraes2000condensed}. In the weak-field approximation \cite{Vilenkin1994}, the linearized line element representing the spacetime of a straight cosmic string oriented along the $z$-axis is given by
\begin{equation}
ds^2=-dt^2+dr^2+ r^2 d\theta^2 +dz^2, \nonumber \label{Vilenkinmetric}
\end{equation}
with $0 < \theta < 2 \pi \alpha$ and $\alpha = 1 - 4G\mu$. This geometry is globally conical, where the conical deficit angle is $4 \pi G\mu$, with $G$ being the Newton's constant and $\mu$ standing for the linear mass density of the cosmic string (in this work, we set $c = 1$). This metric is known as the {\it straight cosmic string} spacetime since it does not account for the effect of small-scale structures, known as kinks or wiggles. In this case, we have an ideal and linear string, which is why the name includes the word {\it straight} on it. On the other hand, if we consider the wiggles, we have the known wiggly cosmic string \cite{Vilenkin1990} that possesses a more significant mass density, yielding a relevant Newtonian potential. The term $G\mu$ associated with a GUT cosmic string is known to be very tiny, namely $10^{-6}$ or $10^{-7}$ \cite{hindmarsh1995cosmic}. However, at the regime of supermassive cosmic strings, the term $G\mu$ can become so much bigger \cite{ortiz1991new, linet1990supermassive, laguna1989spacetime, gonzalez1995extreme}.

Our main aim of this work is to investigate the effect of a cosmic string on a two-dimensional hydrogen atom. In a previous study \cite{de2002non}, the three-dimensional hydrogen atom in the presence of a cosmic string considers the standard ($\propto 1/r$) Coulombian interaction with the cosmic string metric expressed in spherical coordinates rather than cylindrical, as is usual. In contrast, our proposal here presents the hydrogen atom with a logarithmic interaction (2D Coulomb potential) in the presence of a cosmic string while maintaining the original cylindrical/conical symmetry of spacetime's metric. Consequently, the two-dimensional atom consists of an electron and proton attracted to each other by electrical forces produced by the logarithmic potential energy, as is widely recognized \cite{andrew1990hydrogenic, kventsel1981thomas}.

In addition to its pedagogical importance, the two-dimensional hydrogen atom can be considered an excellent analog for some physical systems. For instance, in a very thin semiconductor film, when the semiconductor's size is much smaller than the effective electron's radius, a circumstance in which the electron-hole pair behaves similarly to a 2D hydrogen atom \cite{dasgupta1981comments}. Furthermore, the logarithmic potential form appears in the study of electrons under the influence of a region with a hollow cylinder capacitor and a central straight wire \cite{gesztesy1978electrons, Griffiths:1492149}; and also in the study of light propagation around a wiggly cosmic string, where an optical fiber with a refractive index that mimics the radial wave modes is proposed \cite{azevedo2017wiggly}.

The manuscript solves the eigenvalue problem for the two-dimensional hydrogen atom by employing the Finite Difference Method, which has yet to be previously used for such a purpose. The following sections of the manuscript are organized as follows: Section \ref{2DH} presents the wave equation for the two-dimensional hydrogen atom in the presence of a cosmic string. Later, in Sec. \ref{graphics}, we depict the probability distribution function, and we also show a table with eigenvalues and expected values to demonstrate the influence of the topological factor $\alpha$. Section \ref{analogy} explores the analogy between the two-dimensional hydrogen atom and a two-dimensional monolayer semiconductor. Finally, Sec. \ref{conclusions} concludes the manuscript with our final considerations.

\section{The two-dimensional hydrogen atom}\label{2DH}

In this section, we investigate a hydrogen atom formed by two-dimensional charges, an electron $-e$, and a proton $+e$, in the presence of a straight cosmic string.  The potential energy function that is the solution of the Gauss' law and correctly describes the two-dimensional hydrogen atom is \cite{Eveker1990, Austria1985, PhysRevA.9.2617,Jackson:1998nia}
\begin{equation}
V(r) = \frac{e^2}{2\pi\epsilon_0} \ln \left( \frac{r}{r_0}\right),
\label{Potential}
\end{equation}
where $r$ is the distance separating the two charges, $r_0$ is a scale constant,  and the fundamental constant $\epsilon_0$ is dependent upon on the dimension $[\epsilon_0]=C^2s^2/m^d kg$ (where $d=2 $ for two dimensions) \cite{andrew1990hydrogenic,morales1996analytical}.

To study a non-relativistic particle in curved spacetime, we use the Schrödinger equation in the form
\begin{equation}
\left[-\frac{\hbar^2}{2m}\nabla^2 + \frac{e^2}{2\pi\epsilon_0} \ln \left( \frac{r}{r_0}\right) \right]\psi(r,\theta) = E \psi(r,\theta), \label{waveequation2d}
\end{equation}
where
\begin{equation}
\nabla^2=\frac{1}{\sqrt{-g}}\partial_{ij}(\sqrt{-g}g^{ij}\partial_j),
\end{equation}
with $i,j=1,2,3$, $g=\det(g_{ij})$ \cite{dowker1974covariant} and $m=m_{e} m_{p}/(m_{e}+m_{p})$ is the reduced two-dimensional mass of the system electron-proton.

By setting $dz=0=z$ in the cosmic string metric (planar motion) and substituting Eq. \eqref{Potential} into Eq. (\ref{waveequation2d}), we get the Schr\"{o}dinger equation 
\begin{align}
&-\frac{\hbar^2}{2m }\left( \frac{\partial^2 }{\partial r^2}+\frac{1}{r}\frac{\partial }{\partial r}+\frac{1}{r^2}\frac{\partial^2 }{\partial \theta^2}\right)\psi +\frac{e^2}{2\pi\epsilon_0}\ln \left(\frac{r}{r_0}\right) \psi \notag \\& = E \psi. \label{change0}
\end{align}

Due to the new range of $\theta$ (namely $0 < \theta < 2 \pi \alpha$) and the periodicity condition given by
\begin{eqnarray}
\psi(r,\theta)=\psi(r,\theta+2 \pi \alpha),
\label{p}
\end{eqnarray}
it leads to the quantization of the angular momentum, whose eigenvalues are now defined according to the topological charge $\alpha$, which characterizes the conical geometry. This way, the solution of the wave equation becomes expressed in the form
\begin{equation}
\psi(r,\theta)=e^{i {l}_{\alpha} \theta}f(r), \label{ansatz}
\end{equation}
where ${l}_{\alpha}$ represents the angular momentum quantum number defined by ${l}_{\alpha}={l}/{\alpha} \in \mathds{Z}$, with $l=0, \pm 1, \pm 2, \pm 3, \dots$ \cite{marques2006comment}.

The quantization of the angular momentum imposes a selection rule on the allowed values of $\alpha$ itself, which can now only be a proper fraction, i.e.
\begin{equation}
\alpha=\frac{p}{q},\quad p<q,
\end{equation}
thus, the angular momentum quantum number reads
\begin{equation}
l_\alpha=\frac{q}{p}l,
\end{equation}
being it an integer number, we guarantee it by setting
\begin{equation}
l=kp, \quad k\in \mathds{Z},
\end{equation}
and, consequently, the respective quantum numbers for the angular momentum becomes
\begin{equation}
l_\alpha=kq.
\end{equation}
We quickly observe that for two distinct values $\alpha_1$ and  $\alpha_2$, there exists a common set of values $l_1=l_2$ which generate distinct values of angular momentum quantum numbers, i.e.,  $l_{\alpha_1}\neq l_{\alpha_2}$.
Below, Table \ref{table1} shows some values of $\alpha$ and the respective values for $l$ and $l_\alpha$. It is important to note that for a given value of $l$, only some $\alpha$'s values are allowed, leading to different values of $l_\alpha$. This attribute is essential, and we will discuss it further.

Similarly, by considering the proper fraction $\alpha_c=1-\alpha$, the respective allowed values for $l$ and $l_{\alpha_c}$ are
\begin{eqnarray}
l&=&k(q-p), \quad k\in \mathds{Z}, \\[0.2cm]
l_{\alpha_c}&=&kq.
\end{eqnarray}
Then, the momentum angular quantum numbers are precisely the same for the complement proper fraction.

From equation \eqref{change0}, we must find the eigenvalues $E$ and the corresponding eigenfunctions $\psi$ both accounting the effects of the topological factor $\alpha$, with the eigenfunctions obeying the normalization condition
\begin{equation}
\int_{0}^{2\pi}  \int_{0}^{\infty}  |\psi(r,\theta)|^{2} \, r dr d\theta =1.
\end{equation}

By substituting the ansatz \eqref{ansatz} into (\ref{change0}), we attain the radial equation:
\begin{align}
&-\frac{\hbar^2}{2m }\left(\frac{d^2 }{dr^2}+\frac{1}{r}\frac{d}{d r}-\frac{{l}_{\alpha}^2}{ r^2}\right)f(r)  +{\frac{e^2}{2\pi\epsilon_0}}\ln \left(\frac{r}{r_0}\right) f(r)\notag \\& = E_\alpha f(r). \label{change1}
\end{align}
The $\alpha$ in the energy $E_\alpha(=E)$ indicates that the energy states of the 2D hydrogen atom now explicitly depend on the defect geometry.

By using a suitable dimensionless variable through the transformation
\begin{equation}
r \to   \left(\frac{2\pi\epsilon_0 \hbar^2 }{2m e^2}\right )^{-1/2} r,
\end{equation}
we obtain the dimensionless radial equation
\begin{align}
&-\frac{d^2 f(r)}{dr^2}-\frac{1}{r}\frac{df(r)}{d r}+\left(\frac{{l}_{\alpha}^2}{ r^2}  +\ln {r} \right) f(r)\notag \\&= E_{n , {l}_{\alpha}} f(r) , \label{change12DH}
\end{align}
where the eigenvalues $E_{n, {l}_{\alpha}}$ depending on the quantum numbers $n$ (radial quantum number) and $l_\alpha$ (the angular momentum one). The term $E_{n, {l}_{\alpha}}$ plays the role of the dimensionless energy related to $E_\alpha$ as,
\begin{equation}
E_\alpha=\frac{e^2}{2 \pi \epsilon_0} \left[E_{n , {l}_{\alpha}}+\frac{1}{2} \ln\left(\frac{2 \pi \epsilon_0 \hbar^2}{2m e^2 r_0^2}\right)\right].    \label{energy2DH}
\end{equation}
This equation resembles Eq. (8) obtained in Refs. \cite{PhysRevLett.114.107401, PhysRevB.108.155421} (also see Eq. (8) in Ref. \cite{Keldysh1979CoulombII}), which determines the spectrum of two-dimensional excitons with logarithmic interaction in semiconductor materials. The study of these 2D excitons is an intriguing topic, and we will dedicate Section \ref{analogy} to explore their analogies with the fundamental aspects of the ones obtained in the model we are presenting here.

\begin{figure}[!t]
\centering
\includegraphics[scale=0.8]{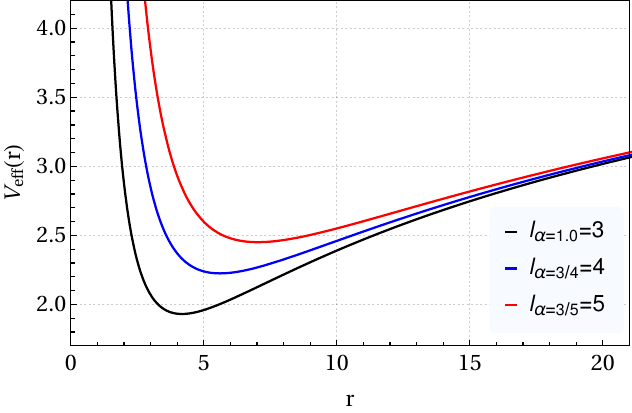}
\caption{The effective potential for a fixed $l = 3$ modified by the cosmic string parameter $\alpha$. According to Table \ref{table1}, the blue and red curves represent, respectively, both $l_{\alpha=3/4}=4$ and $l_{\alpha=3/5}=5$. The black line is the case of $l_{\alpha=1}=l=3$.}
\label{potential}
\end{figure}

Let us make the change of variables $f(r)=r^{-1/2}R(r)$, such that the equation \eqref{change12DH} becomes
\begin{align}
&-\frac{d^2R(r)}{d r^2}+V_{eff}(r) R(r)=  E_{n , {l}_{\alpha}} R(r), \label{change}
\end{align}
where $V_{eff}(r)$ represents the effective potential given by
\begin{equation}
V_{eff}(r)=\frac{l_{\alpha}^2-{1}/{4}}{  r^2}+\ln {r}, \label{pot}
\end{equation}
which, by the way, only supports bound states.
We also observe that for a fixed $l$, decreasing the $\alpha$'s values makes the effective potential less deep, whereas the global minimum moves away from the origin, as shown in Fig. \ref{potential}.

Considering $\alpha=1$ (Euclidean space), we recover from Eq. \eqref{change} the radial Schr\"{o}dinger equation for the two-dimensional hydrogen atom without curvature, as given in Refs. \cite{Eveker1990,Austria1985}. An exact analogous wave equation can be found in Refs. \cite{PhysRevLett.114.107401,PhysRevB.108.155421}.  In the following, as shown in Table \ref{table2}, we present the spectrum
obtained from the above references and compare it with our numerical analysis.
Furthermore, the equation (\ref{change}) is also analogous to the radial wave equation found in the study of massless and massive field propagating in a wiggly cosmic string waveguide \cite{azevedo2017wiggly}. On the other hand, here, the wave equation (\ref{change}) describes the interaction between a proton located at the vertex of a conical space and an electron outside the cone. In other words, we can also interpret it as describing an electron near a cosmic string. To solve Eq. (\ref{change}), we use a Finite Difference Method \cite{van1996, Franklin2011, azevedo2017wiggly}, which enables us to conduct numerical computations for determining the eigenvalues and eigenfunctions of the system. Thus, the states are then characterized by the quantum numbers $n$ (radial quantum number) and $l_\alpha$ (the angular momentum one), as we will further discuss in the following. To our knowledge, it is the first time that the Schr\"{o}dinger equation with a logarithmic potential is solved with such a method.
\begin{table}[H]
\centering
\begin{tabular}{c@{\hspace{1.2cm}}c@{\hspace{1.2cm}}c}\\
\toprule
 $l$ & $\alpha$ & $l_\alpha$ \\
 \midrule
 0 & $1/2<\alpha  \le 1^*$ & 0 \\
 1 & 1 & 1 \\
 2 & 1 & 2 \\
 3 & 3/5 & 5 \\
 3 & 3/4 & 4 \\
 3 &  1 & 3 \\
 4 & 4/5 & 5 \\
 4 & 1 & 4 \\
 5 & 5/8 & 8 \\
 5 & 1 & 5 \\
 6 & 3/5 & 10 \\
 6 & 3/4 & 8 \\
 6 & 1 & 6 \\
 7 & 7/10 & 10 \\
 7 & 7/8 & 8 \\
 7 &  1 & 7 \\
 8 & 4/5 & 10 \\
 8 & 1 & 8 \\
 9 & 9/16 & 16 \\
 9 & 3/5 & 15 \\
 9 & 3/4 &  12 \\
 9 & 9/10 & 10 \\
 9 &  1 & 9 \\
 10 & 5/8 & 16 \\
 10 &  1 & 10 \\
\toprule
\end{tabular}
\caption{Table with some allowed values of $l_\alpha$ for $1/2<\alpha  \le 1$. A larger array of possibilities can be displayed for larger values of $l$ and its corresponding value of $\alpha$. One characteristic observed reveals that $\alpha$ is a proper fraction when $\alpha\neq 1$. $^*$For $l=0$, all values of $\alpha$ leads to the same $l_\alpha=0$.}
 \label{table1}
\end{table}

\begin{figure}[!t]
\centering
\includegraphics[scale=0.5]{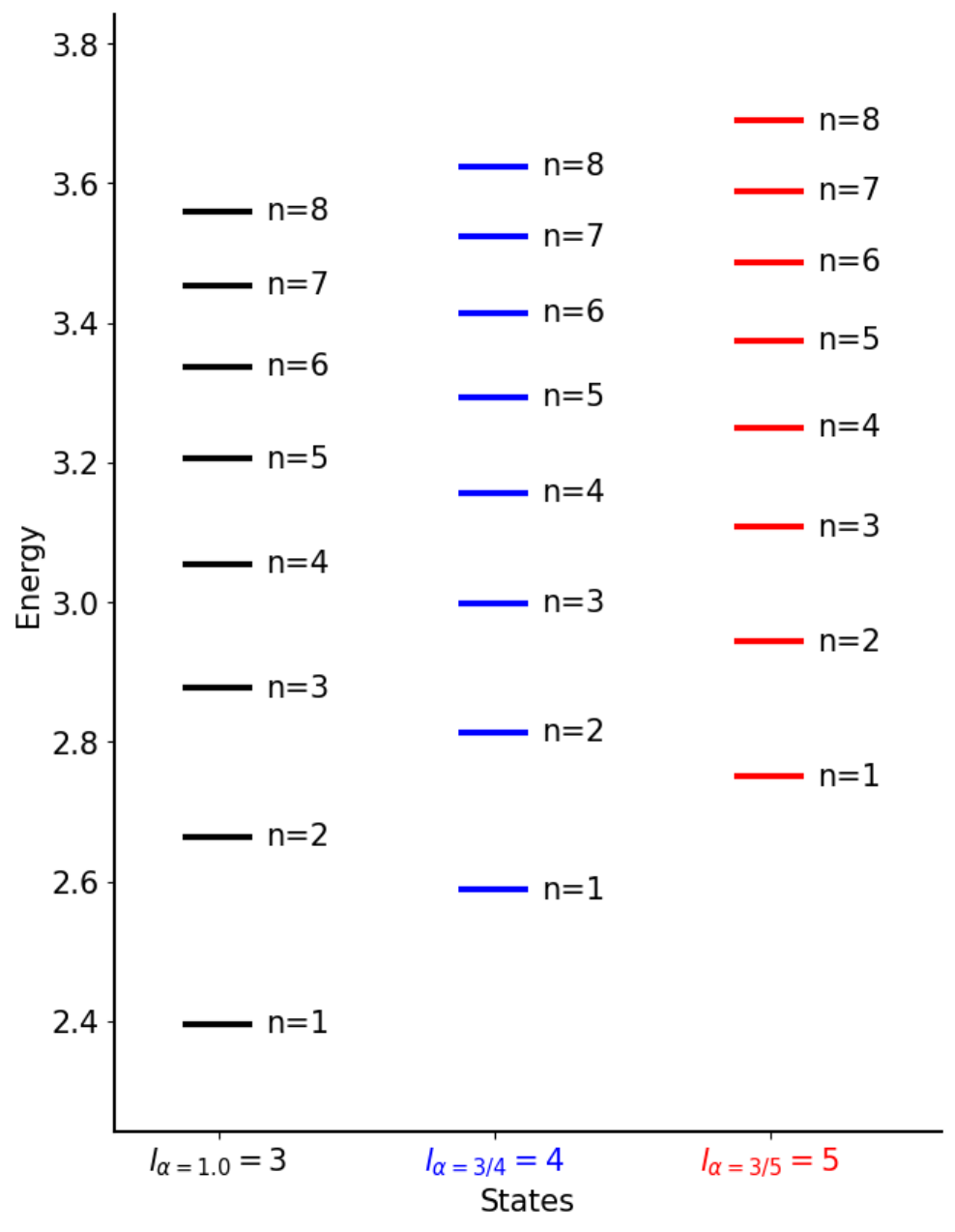}
\caption{The energy level diagram for a fixed $l = 3$ influenced by the cosmic string factor $\alpha$, which causes upper shifts in the energies values. The energy levels sate in blue and red serve for $l_{\alpha=3/4}=4$ and $l_{\alpha=3/5}=5$ (the black lines depict the ones for $\alpha=1$), respectively.}
\label{diag}
\end{figure}

\begin{table*}[]
\centering
\begin{tabular}{c@{\hspace{0.75cm}}c@{\hspace{0.75cm}}c@{\hspace{0.8cm}}c@{\hspace{0.8cm}}c@{\hspace{0.8cm}}c@{\hspace{0.8cm}}c@{\hspace{0.8cm}}c@{\hspace{0.75cm}}c}
\toprule
$l$ &  $\alpha$ &  $l_\alpha$  &  {$E_{n, {l}_{\alpha}}$ (This work)} &   $E_{n, {l}}$ (Ref. \cite{Austria1985}) & $E_{n, {l}}$ (Ref. \cite{PhysRevLett.114.107401}) & $\braket{\psi|r|\psi}$ & $\braket{\psi|r^2|\psi}$ &$n$  \\
\midrule
& & & 0.52639  & 0.52664  & 0.5265 &1.72478  &
3.70635&1  \\
& & & 1.66121 & 1.66134  & 1.661 &4.41729  &21.35860& 2 \\
0 & $1/2<\alpha\le 1$ & 0 & 2.17715 &2.17724 &2.177&   7.27366 &56.78452& 3 \\
& & & 2.51543 & 2.51553  & $\cdots$& 10.15265 &109.99873& 4  \\
& & & 2.76761 & 2.776   & $\cdots$ & 13.03889  &180.99504&5 \\ \\
& & & 1.38618 & 1.38621  &1.386& 2.67106  & 8.41683& 1 \\
& & & 2.00947 & 2.00951   &2.009 &5.16206 &31.11762& 2 \\
1 & 1 & 1 & 2.39434 & 2.39437  &-- &7.65924  &68.21615& 3 \\
& & & 2.67267 & 2.6727  &$\cdots$&10.15930   &119.76373&4  \\
& &  & 2.89061 & $\cdots$  &$\cdots$ &12.66100  &185.78132& 5 \\ \\
& & & 1.84437 & 1.84440  & 1.844 &4.08285 &18.65697& 1 \\
& & & 2.27586 & 2.27645   & $\cdots$ & 6.57581 &49.77528& 2 \\
2 & 1 & 2 & 2.58005 & 2.58021  & $\cdots$ & 9.07092&95.20211& 3  \\
& & & 2.81447 & $\cdots$ &  $\cdots$& 11.56805 &155.01022& 4 \\
& & & 3.00496 & $\cdots$ & $\cdots$& 14.06673 &229.23714 & 5 \\ \\
& & & 2.15785 & 2.15785  & $\cdots$ &5.49590  & 32.89834& 1 \\
& & & 2.48812 & 2.48812  & $\cdots$ &7.99017  &72.46662&  2 \\
3 & 1 & 3 & 2.73905 & $\cdots$ & $\cdots$ &10.48483&126.28602& 3 \\
& & & 2.94097 & $\cdots$ &$\cdots$ &12.98064 &194.43390& 4 \\
& & & 3.10968 & $\cdots$ &$\cdots$  &15.47763  &276.95583& 5 \\ \\
\begin{tabular}{@{}c@{}}3\\4\end{tabular} & \begin{tabular}{@{}c@{}}3/4\\1\end{tabular}& 4& \begin{tabular}{@{}c@{}}2.39628\\2.66388\\2.87727\\3.05438\\3.20558\end{tabular} & \begin{tabular}{@{}c@{}}2.396302\\$\cdots$\\$\cdots$\\$\cdots$\\$\cdots$\end{tabular} &
\begin{tabular}{@{}c@{}}$\cdots$ \\
$\cdots$  \\
$\cdots$ \\
$\cdots$  \\
$\cdots$  \\
\end{tabular} &
\begin{tabular}{@{}c@{}}6.90943 \\
9.40462  \\
11.89934  \\
14.39459  \\
16.89063  \\
\end{tabular} &
\begin{tabular}{@{}c@{}}51.14022\\
99.17210\\
161.41547\\
237.94640\\
328.81377\\
\end{tabular} &
\begin{tabular}{@{}c@{}}1\\2\\3\\4\\5\end{tabular} \\
\\
\begin{tabular}{@{}c@{}}3\\4\\5\end{tabular} & \begin{tabular}{@{}c@{}}3/5\\4/5\\1\end{tabular}& 5& \begin{tabular}{@{}c@{}}
2.58872\\2.81367\\2.99922\\3.15686\\3.29374\end{tabular} & \begin{tabular}{@{}c@{}}$\cdots$\\$\cdots$\\$\cdots$\\$\cdots$\\$\cdots$\end{tabular} &
\begin{tabular}{@{}c@{}}$\cdots$ \\
$\cdots$  \\
$\cdots$ \\
$\cdots$  \\
$\cdots$  \\
\end{tabular} &
\begin{tabular}{@{}c@{}}8.32320  \\
10.81905  \\
13.31401  \\
15.80906  \\
18.30456\end{tabular} &
\begin{tabular}{@{}c@{}}73.38236 \\
129.88485\\
200.56992\\
285.51028\\
384.75403\end{tabular} &
\begin{tabular}{@{}c@{}}1\\2\\3\\4\\5
\end{tabular} \\
\toprule
\end{tabular}
\caption{Energy levels, expected values $\braket{\psi|r|\psi}$ and $\braket{\psi|r^2|\psi}$ for different values of $l$, $\alpha$ and $l_\alpha$. The changes due to the presence of the parameter $\alpha$ are observed clearly in the energy levels and the calculation of expected values. The table also includes energy states calculated in other environments that support wave equations analogous to Eq. (\ref{change}).}
 \label{table2}
\end{table*}

\subsection*{Numerical Analyses}
\label{graphics}

For the numerical analyses, we use suitable values of linear mass densities to obtain a general aspect of the effects of cosmic string geometries on the spectrum and eigenfunction of the two-dimensional hydrogen atom. In this way, we choose $1/2 < \alpha  \le 1$ (with $\alpha=1$ being the Euclidean limit).
We adopted the physical constants, such as Planck's constant and proton and electron masses, in the SI units.

In Table \ref{table1}, we have a list of the first allowed values of $l$ and the $\alpha$'s values that combine to give an integer ${l}_{\alpha}$, clearly showing how the $\alpha$ affects the angular momentum quantum number. Each value of $\alpha$ generates different values for ${l}_{\alpha}$. On the other hand, for the same value of $l$, there exist distinct $\alpha$ values. Thus, given the defect's presence, its essential feature is to select the corresponding values of $l$, engendering the integer values for ${l}_{\alpha}$. Such a situation contrasts with the planar case ($\alpha=1$). Consequently, the most important characteristic of the presence of the topological defect is the quantum numbers accessed by the angular momentum $l_\alpha=l/\alpha$, which can alter the hydrogen atom's energy states, eigenfunctions, and expected values.

\begin{figure}[!t]
\centering
\includegraphics[scale=0.8]{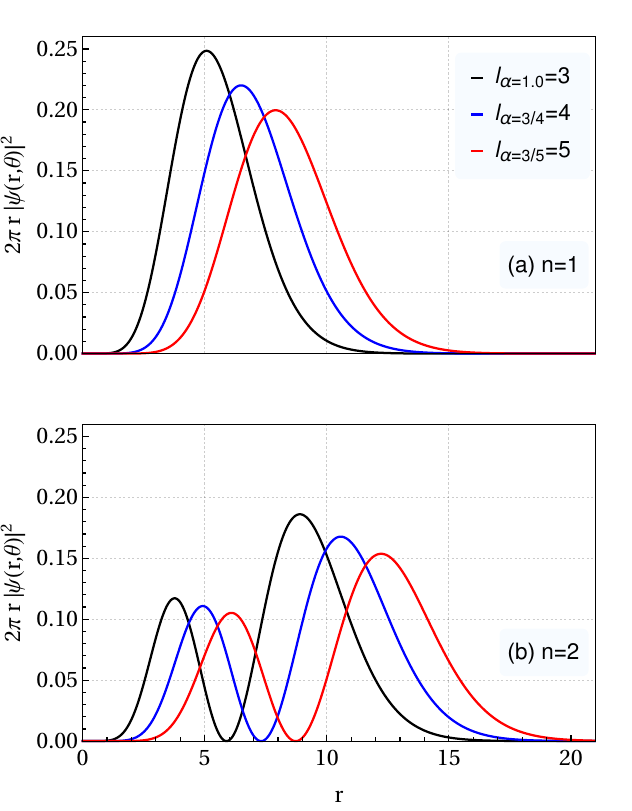}
\caption{The probability distribution function for a fixed $l = 3$ is influenced by the cosmic string factor $\alpha$, which causes shifts in probability values. In (a) for radial quantum number $n=1$ and in (b) for radial quantum number $n=2$.  The black curve is for $\alpha=1$, the blue and red curves represent, respectively, both $l_{\alpha=3/4}=4$ and $l_{\alpha=3/5}=5$.}
\label{prob1}
\end{figure}

\begin{figure}[!t]
\centering
\includegraphics[scale=0.32]{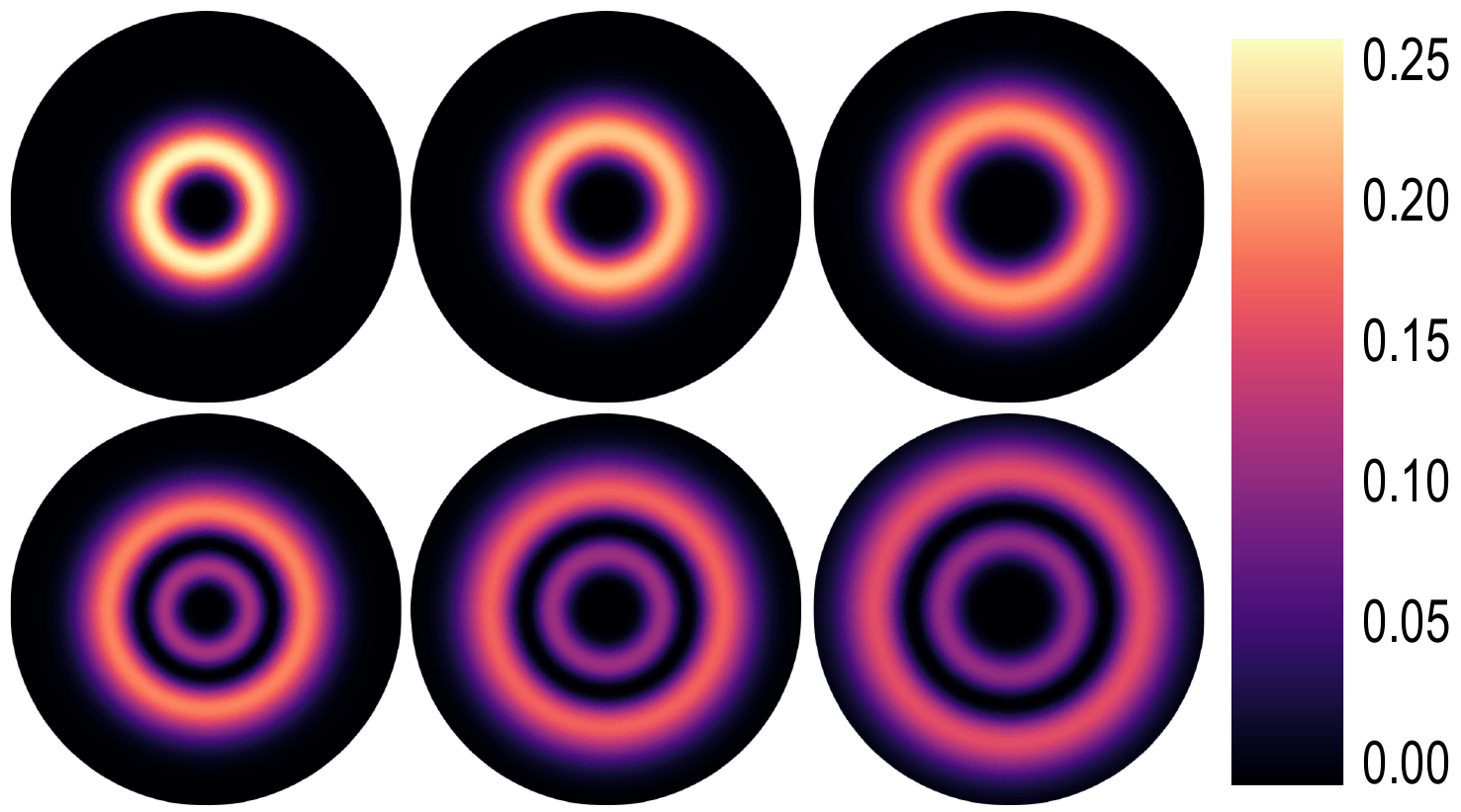}
\caption{The disks of probability indicating regions with higher probabilities (higher intensities). We use the same $l=3$ and $n$ as done in Fig. \ref{prob1}. The disks are described as follows: the first line is for $n=1$, and the second line is for $n=2$. First column for $l_{\alpha=1.0}=3$, second for $l_{\alpha=3/4}=4$ and third for $l_{\alpha=3/5}=5$.}
\label{prob2}
\end{figure}

Table \ref{table2} shows the energy values and expected values for the corresponding $l$ and $\alpha$. The Finite Difference Method results perfectly match the linear variational technique used by Asturias and Aragón \cite{Austria1985} to solve the 2D hydrogen atom in the absence of the topological defect. Furthermore, the energy levels match the spectrum provided in Ref. \cite{PhysRevLett.114.107401} in the treating two-dimensional excitons in semiconductors. The following section will explore the two-dimensional excitons with logarithmic approximation and present new results for the semiconductor's energy states in SI units. By looking at the table, we can detect that the value of energy increases with the quantum numbers $n$ and $l$. We also notice that introducing the defect does change the energy eigenvalue significantly for the eigenstates; see the diagram in Fig. \ref{diag}. Our choices for $l_{\alpha}$ in Figs. \ref{potential} and \ref{diag} (and later in Figs. \ref{prob1} and \ref{prob2}) are because those represent the first values allowed for  $\alpha$ besides the Euclidean value $\alpha=1$.
Once we find the eigenfunctions, we can calculate the probability distributions. In this way, Figure \ref{prob1} presents the probability distribution functions for fixed $l=3$ and three values of the topological factor: $\alpha=1$ (flat space), $\alpha=3/4$, and $\alpha=3/5$. In Fig. \ref{prob1}(a), for $ n=1$, we can observe that as the value of $l_{\alpha}$ increases, the density probability's maximum value decreases while becoming more distant from the origin. Fig. \ref{prob1}(b) shows a similar effect but for a more significant radial quantum number, $n=2$. In Fig. \ref{prob2}, we have a disk of probabilities representing the same effect in a more visually appealing manner.  We see that the size of the rings increases with $l_{\alpha}$. Furthermore, the ring numbers are proportional to the radial quantum number $n$.
We now calculate the expected values: $\langle \psi|r|\psi \rangle$ and $\langle \psi|r^2|\psi \rangle$. The results clearly demonstrate the effect of the topological defect, leading to an increase in the expected values, as shown in Table \ref{table2}. For instance, the expected value $\langle \psi|r|\psi  \rangle$ for $l=3$ and $n=1$ (with $\alpha=1$, $\alpha=3/4$, and $\alpha=3/5$) is strongly determined by the value of $r$ where the probability distribution function has a peak in Fig. \ref{prob1}, as expected. Using these expected values, we can calculate optical properties such as the oscillator strength (consequently, absorption coefficients and refractive index changes) \cite{PhysRevB.98.125406, https://doi.org/10.1002/pssb.202200097} and diamagnetic shift  \cite{doi:10.1021/acs.nanolett.6b03276, stier2016exciton,PhysRevB.97.195408, PhysRevB.57.9088, PhysRevB.107.L201407, PhysRevLett.120.057405}.

\section{Rytova-Keldysh potential and logarithmic approximation}
\label{analogy}

We have mentioned an analogy between semiconductors and the hydrogen atom. For instance, in the Wannier-Mott exciton model \cite{PhysRev.52.191}, the 3D Coulomb potential rules the interaction between excitons (electron-hole pairs) in three-dimensional semiconductors, where the hole assumes a role similar to that of a proton \cite{ZAKHARCHENYA2005171, PhysRevLett.116.056401}. However, the analogy regarding exciton binding energy in two-dimensional semiconductors is not direct. While the 2D excitons in higher energy states exhibit 3D hydrogen characteristics, those in lower energy states do not display the 3D hydrogen behaviors \cite{PhysRevLett.113.076802}. The 2D exciton in a plane semiconductor represents an electron-hole pair confined within a thin monolayer dielectric whose interaction is known as the Rytova-Keldysh (RK) potential \cite{Rytova,Keldysh1979CoulombII}
\begin{equation}
 V_{RK}(r)= -\frac{e^2}{4\pi  \epsilon_0 \kappa} \frac{\pi}{2 r_{s}}\left[  H_{0}\left(\frac{r}{r_{s}}\right) - Y_{0}\left(\frac{r}{r_{s}}\right)   \right] ,
 \label{RKpot}
\end{equation}
where $H_0$ is the Struve function and $Y_0$ the Bessel function of the second kind; $\kappa$ is the dielectric constant of the surrounding environment, and $r_{s}=2\pi\alpha_s/\kappa$ is the screening length with $\alpha_s$ being the polarizability of the 2D material \cite{PhysRevB.97.195408}. This potential exhibits behavior similar to a 3D Coulomb potential at long distances but diverges logarithmically at short distances. Neglecting the logarithmic contribution will lead to unreliable results for the excitons' ground state description, resulting in noticeable differences in the Rydberg series, wavefunctions, and binding energies of exciton states \cite{PhysRevLett.120.057405,stier2016exciton,PhysRevLett.113.076802, PhysRevLett.113.026803}.

\begin{figure}[!t]
\centering
\includegraphics[scale=0.8]{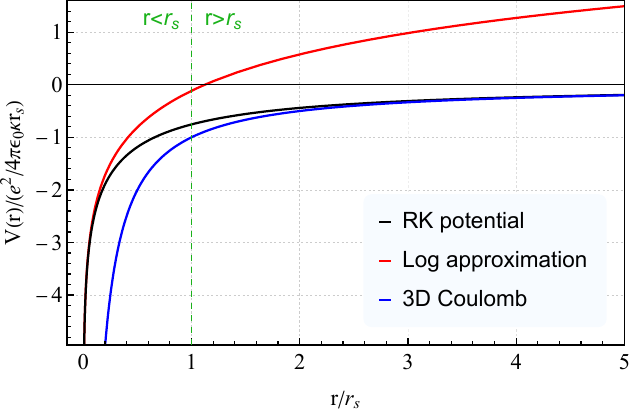}
\caption{Comparison between the logarithmic approximation given by Eq. \eqref{pot-r0}, the RK potential described in Eq. \eqref{RKpot}, and the 3D Coulomb potential from Eq. \eqref{coulomb3d}.}
\label{potentialS}
\end{figure}

Recognizing the absence of a complete analytical solution for the problem, some authors \cite{PhysRevB.105.L201407,PhysRevB.84.085406,PhysRevB.102.125303} (also see \cite{PhysRevB.108.155421,PhysRevLett.114.107401}) have proposed approximate solutions. In these approaches, the RK interaction is given in the form
\begin{equation}
    V(r)\approx-\frac{e^2}{4 \pi \epsilon_0 \kappa}\frac{1}{r}
    \label{coulomb3d}
\end{equation}
for $ r\gg r_{s}$ and in the logarithmic approximation
\begin{equation}
    V(r)\approx \frac{e^2}{4 \pi \epsilon_0 \kappa } \frac{1}{r_{s}}\left[\ln\left( \frac{ r}{2 r_{s}}\right) + \gamma \right]
    \label{pot-r0}
\end{equation}
for $ r\ll r_{s}$ \cite{PhysRevB.84.085406}
(corresponding to the cases of null $r_{s}\to 0$ and very large screening lengths $r_{s}\to \infty$, respectively \cite{PhysRevB.97.195408,kirichenko2021influence}), $\gamma \approx 0.5772 $ is the Euler constant. Although the equations (\ref{coulomb3d}) and (\ref {pot-r0}), strictly speaking, are derived for values of $r$ much greater and much smaller than $r_{s}$, Figure 1 in ref. \cite{PhysRevB.91.245421} (also see Fig. \ref{potentialS}, in this work) demonstrates that the logarithmic approximation of the RK potential (\ref{RKpot}) is already accurate for sufficiently small values of $r/r_{s}$. As mentioned earlier, a complete analytical solution to the Schr\"odinger equation for a logarithmic potential has not been found yet. However, asymptotic solutions for the logarithmic potential in the second limit region mentioned above ($\propto \ln r$) have been reported in the literature. In Section \ref{2DH}, we numerically solved the Schrödinger equation in polar coordinates with the logarithmic potential. The results obtained in that Section will be utilized here.

Following Nguyen-Truong's work \cite{PhysRevB.105.L201407}, which approximately solved the Schrödinger equation for 2D Wannier-Mott excitons in the effective-mass approximation, we will present a complete numerical solution by considering the same starting point given by the equation
\begin{equation}
\left(-\frac{\hbar^2}{2\mu }\nabla^2 + V(r)\right)\psi(r,\theta)=E \psi(r,\theta),\label{waveequationS}
\end{equation}
where $\mu$ represents the exciton reduced mass, $\nabla^2$ is the Laplacian in ordinary polar coordinates (Euclidean spacetime), and the interaction $V(r)$ is as defined in Eq. (\ref{pot-r0}). Namely, in Ref. \cite{PhysRevB.105.L201407}, an interaction with a logarithmic approximation (satisfying both limits in Eqs. \eqref{coulomb3d} and \eqref{pot-r0}) was applied to solve Eq. (\ref{waveequationS}). However, another approximation valid for large $r$ was later employed to derive the energy expression.

We start by writing the wave function in polar coordinates as
\begin{equation}
\psi(r,\theta)=e^{i {l} \theta}f(r), \label{ansatzS}
\end{equation}
where $l=0, \pm 1, \pm 2 , \pm 3, \dots$ is the quantum angular momentum. In the region $r\ll r_{s}$, the equation (\ref{waveequationS}) reads
\begin{align}
-\frac{\hbar^2}{2\mu }&\left(\frac{d^2 }{dr^2}+\frac{1}{r}\frac{d}{d r}-\frac{{l}^2}{r^2}\right)f(r) \notag \\ & +\frac{e^2}{4 \pi \epsilon_0 \kappa } \frac{1}{r_{s}}\left[\ln\left( \frac{ r}{2 r_{s}}\right)+ \gamma \right] f(r)= E f(r). \label{change1S}
\end{align}
Considering a dimensionless radial coordinate through the variable transformation,
\begin{equation}
r \to   \left(\frac{4\pi\epsilon_0 \kappa r_{s}\hbar^2 }{2\mu e^2}\right )^{-1/2} r,
\end{equation}
we achieve the dimensionless radial equation
\begin{equation}
-\frac{d^2 f(r)}{dr^2}-\frac{1}{r}\frac{df(r)}{d r}+\left(\frac{{l}}{ r^2}  +\ln {r} \right) f(r)= E_{n , {l}} f(r) , \label{change1x}
\end{equation}
where the eigenvalues $E_{n, {l}}$ depending on the quantum numbers $n$ and $l$ and it is related to $E$ through
\begin{equation}
E = \frac{e^2}{4 \pi \epsilon_0 \kappa } \frac{1}{r_{s}} \left[(E_{n , {l}}+ \gamma)+\frac{1}{2} \ln\left(\frac{ 4\pi\epsilon_0 \kappa\hbar^2}{2\mu e^2 4 r_{s}}\right) \right].
\label{energy}
\end{equation}
This equation resembles Eq. (21) analyzed in Ref. \cite{PhysRevB.91.245421}, where a variational analytical method was employed to study 1s excitons in two-dimensional crystals with potential Eq. \eqref{pot-r0}. Specifically, they would be mathematically identical if the sum $E_{n, {l=0}}+\gamma$ will equal $3/2$. However, here our computations show that $E_{n, {l=0}}+\gamma$ has the value $1.104$.

By making the change of variables $f(r)=r^{-1/2}R(r)$, the equation \eqref{change1x} can be written as
\begin{align}
&-\frac{d^2R(r)}{d r^2}+\left(\frac{l^2-{1}/{4}}{  r^2}+\ln {r}\right)R(r)=  E_{n , {l}} R(r). \label{changeS}
\end{align}
Equation (\ref{changeS}) is mathematically similar to Eq. (\ref{change}) for the 2D hydrogen atom with $\alpha=1$ (absence of the topological defect). This similarity is because the potential given by Eq. \eqref{pot-r0} closely resembles the 2D hydrogen potential described in Eq. \eqref{Potential}.
Thus, the respective eigenvalues $E_{n,l}$ are also detailed in Table \ref{table2} for the values of $E_{n, l_{\alpha}}$ with $\alpha=1$ (i.e., $E_{n,l}=E_{n, l_{\alpha=1}}$).

\begin{figure}[!t]
\centering
\includegraphics[scale=0.8]{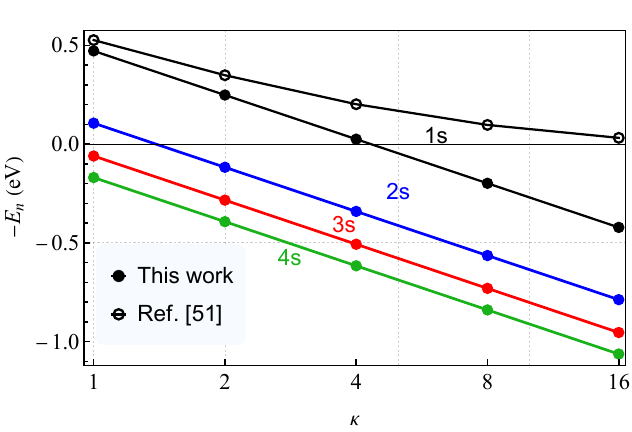}
\caption{The exciton energy levels of the ns-states (ns-excitons); $n=1,2,3,4$ and $l=0$ as a function of the average dielectric constant for MoS${_2}$ are plotted. The 1s exciton state is compared to a result available in the literature. The horizontal axis is on a logarithmic scale.}
\label{energy-semiconductor}
\end{figure}

Table \ref{table3} exhibits the exciton energies (in SI units) found for some semiconductors by using the equation \eqref{energy} in the ground-state $n=1$ and $l=0$ (1s exciton). Our analysis encompasses pure materials and those positioned on a silica (SiO$_2$) substrate. The table shows that the energy value ($-E$(eV)) decreases when the material is on the substrate. This outcome agrees with Refs. \cite{PhysRevB.105.L201407, PhysRevB.97.195408, PhysRevB.92.205418}, where energy decreases upon substrate placement as well; nevertheless, the energy values achieved here slightly differ from the ones in Ref. \cite{PhysRevB.92.205418}. However, this variance happens because the semiconductor analysis occurs within the region  $ r \ll r_{s}$ where the potential is given by Eq. \eqref{pot-r0}. Fig. \ref{energy-semiconductor} shows that our computations closely approximate the energy values attained in the Ref. \cite{PhysRevB.92.205418} for the ground state when $\kappa$ has small values. Besides, the energy sign reverses because the logarithmic approximation becomes less accurate for $r>r_s$; see Fig. \ref{potentialS}. Beyond that, Ref. \cite{PhysRevB.92.205418} notes that the RK potential also becomes questionable for large values of $\kappa$.
 
\begin{table}[!t]
	\centering
	\begin{tabular}{c@{\hspace{0.35cm}}c@{\hspace{0.35cm}}c@{\hspace{0.35cm}}c@{\hspace{0.12cm}}} 
		\toprule
		 Materials & Substrates & This work & Ref. \cite{PhysRevB.92.205418} \\
		\midrule
		\multirow{2}{*}{MoS${_2}$} & Isolated & 0.47164 & 0.5265 \\
		& SiO$_{2}$ & 0.24828 &0.3486 \\
		\multirow{2}{*}{MoSe${_2}$} & Isolated & 0.43685 & 0.4769 \\
		& SiO$_{2}$ & 0.24901 & 0.3229 \\
		\multirow{2}{*}{WS${_2}$} & Isolated & 0.43252 & 0.5098 \\
		& SiO$_{2}$ & 0.18403 & 0.3229 \\
		\multirow{2}{*}{WSe${_2}$} & Isolated & 0.39763 & 0.4564 \\
		& SiO$_{2}$ & 0.18780 & 0.2946 \\
		\toprule
	\end{tabular}
\caption{Table with the negative of the exciton energy $-E_n$(eV) of the semiconductor given from Eq. \eqref{energy} in SI units for the ground-state with $n=1$ and $l=0$. The material's units are sourced from Table I in Ref. \cite{PhysRevB.92.205418}
(for more details, see also the supplemental material for this reference). The isolated material in a vacuum is defined for $\kappa=1$, and when placed in silica (SiO$_{2}$), it is characterized by $\kappa=2$.}
 \label{table3}
\end{table}
The analogy between the 2D hydrogen atom around a cosmic string and a 2D semiconductor is only direct for $l=0$ and $\alpha=1$ (the 2D Euclidean space). Hence, the results in Section \ref{graphics} are valid for describing the 2D semiconductor when considering the previous conditions. The analogy would be even more complete if we consider the semiconductor counterpart analog of the factor $\alpha$, which would change the angular momentum. 
It is well-known that cosmic strings are analogous to disclinations in liquid crystals \cite{moraes2000condensed}. Furthermore, some theoretical models with a disclination-type topological defect have been proposed to study 2D semiconductor quantum rings \cite{PE.2019.109.59, AdP.2019.531.1900254, PE.2021.132.114760, AdP.2023.535.2200371} (also see Ref. \cite{ACS.2018.12.2580}). From a geometric standpoint, the conical symmetry in a semiconductor monolayer may originate from Volterra’s well-known cut and glue process, which removes a wedge of the material and glues the straight edges of the remaining part. On the other hand, it is noteworthy that the analyses and the solutions in the literature often concentrate solely on s-state excitons (with $l=0$) (e.g., \cite{PhysRevB.105.L201407}). This preference comes because the s-state excitons offer a more straightforward path to analytical and approximate solutions and are more experimentally accessible. Therefore, the analogy drawn here holds significant strength.

To conclude, we call attention to the fact that if the equation \eqref{waveequationS} incorporates the cosmic string metric (precisely like the approach in Section \ref{2DH}), then we would account for the effects of the defect geometry on the 2D excitons under the influence of the interaction \eqref{pot-r0}. However, we will observe such an effect only for excitons with $l>0$.

\section{Remarks and conclusions\label{conclusions}} 

We here have studied the 2D logarithmic potential in the context of electron-proton interaction in a two-dimensional hydrogen atom, while some other works consider the conventional 3D Coulomb potential when dealing with similar systems \cite{10.1119/1.11670, 10.1063/1.1503868, 10.1119/1.1973790, 10.1119/1.12546, POSZWA2020114247}. We believe such results can be unreliable for the two-dimensional hydrogen atom in the sense of an actual “planar universe.” Instead, those results could represent the electron-proton interaction confined to a plane immersed in 3D, as clarified by some authors in the references mentioned previously. Nevertheless, Yang et al. \cite{PhysRevA.43.1186} pointed out this trick difference remarkably by saying, {\it ``We would like to point out that 2D in the name `2D hydrogen atom' only emphasizes that the motion of the electron around a positive point charge (not a line charge) is constrained in a plane. This system is not 2D in a strict sense that all fields, including electromagnetic fields, photon emission, angular momentum, and spin, are not confined to a plane."}
The results using the 3D Coulomb potential in the context of 2D exciton semiconductors are satisfactory for large distances (higher energy state), where the potential interaction behaves as $\propto 1/r$. However, at small distances, the interaction behavior becomes logarithmic, and the 3D Coulomb potential fails, as illustrated in Fig. \ref{potentialS} and discussed in Section \ref{analogy}. Notwithstanding, through this work, we again address this misunderstanding and encourage new studies to consider the logarithmic interaction in actual 2D systems, especially in the treatment of 2D semiconductors, in which the interaction behavior is logarithmic (at least for short distances).

Our results show the influence of the topological defect on the energy levels and the expectation values of the 2D hydrogen atom. Including the cosmic string makes selection rules for the allowed values of $l$ and the $\alpha$ factor, which is now a proper fraction; both enable us to have integer eigenvalues of the momentum angular, $l_\alpha$. In other words, it also means that, for a given  $\alpha$, not all values of $l$ are allowed. For instance, a GUT cosmic string has a value of $\alpha$ very close to $1$, thus allowing only greater values of $l$ and, consequently, large values of $l_\alpha$, the momentum angular's quantum numbers. Furthermore, we have shown that our numerical solution for the logarithmic potential leads to an interesting analogy with a two-dimensional semiconductor within a specific region of interest, achieved by considering suitable choices of the physical parameters. We point out that the numerical method used in this manuscript can be easily applied to describe excitons in these types of systems.

By the closure of this work and finalization of Section \ref{analogy}, we were pleasantly surprised by the discovery of Ref. \cite{PhysRevB.108.155421} (notably, we also paid attention to Ref. \cite{PhysRevLett.114.107401}). This reference discusses 2D excitons in semiconductors and mentions their connection with the propagation of fields in the wiggly string spacetime. This exciting discovery at the end of our editing process has significantly enhanced the physical motivations of our paper and increased our confidence in the coverage of the analogy we have outlined.

Finally, it is worthwhile to point out that the present study could contribute to a more profound understanding of the influence of topological defects on the behavior of a  charged quantum particle. For instance, the approach found in Section \ref{2DH} could be applied to the RK potential \eqref{RKpot} to investigate the effect of a cosmic string on a 2D exciton in a semiconductor. By the way, we are working on applying the Finite Difference Method to other systems to solve their wave equations. Advances will be reported elsewhere.

\section*{Acknowledgments}

This work was partially supported by the Brazilian agencies CAPES, CNPq, and FAPEMA. E. O. Silva acknowledges the support from the grants CNPq/306308/2022-3, FAPEMA/UNIVERSAL-06395/22. F. S. Azevedo acknowledges CNPq Grant No. 150289/2022-7. R. C. acknowledges the support from the grants CNPq/306724/2019-7, FAPEMA/Universal-01131/17, FAPEMA/Universal-00812/19, and FAPEMA/APP-12299/22. This study was financed in part by the Coordena\c{c}\~{a}o de Aperfei\c{c}oamento de Pessoal de N\'{\i}vel Superior - Brasil (CAPES) - Finance Code 001. We are grateful to Walter Dal'Maz Silva for assisting us in applying the Finite Difference Method in Python.

\bibliographystyle{apsrev4-2}
%

\end{document}